\documentclass[lettersize,journal]{IEEEtran}

\usepackage{color}
\usepackage{tabularray}

\usepackage{amsmath,amsfonts}
\usepackage{amssymb}
\usepackage{algorithmic}
\usepackage{array}
\usepackage[caption=false,font=normalsize,labelfont=sf,textfont=sf]{subfig}
\usepackage{textcomp}
\usepackage{stfloats}
\usepackage{verbatim}
\usepackage{graphicx}
\usepackage{mathrsfs}
\usepackage{CJK}
\usepackage{indentfirst}
\usepackage{cases}
\usepackage{array}
\usepackage{titlesec}
\usepackage{xcolor}

\def\BibTeX{{\rm B\kern-.05em{\sc i\kern-.025em b}\kern-.08em
    T\kern-.1667em\lower.7ex\hbox{E}\kern-.125emX}}
   
\usepackage{balance}

\begin{document}
\title{Learning-to-solve unit commitment based on few-shot physics-guided spatial-temporal graph convolution network}
\author{Mei Yang, Gao Qiu, Junyong Liu, \emph{Member, IEEE},  Kai Liu, \emph{Senior Member, IEEE}
\vspace{-4ex}



}

\markboth{Journal of \LaTeX\ Class Files,~Vol.~18, No.~9, September~2020}%
{How to Use the IEEEtran \LaTeX \ Templates}
\maketitle
\begin{abstract}
This letter proposes a few-shot physics-guided spatial temporal graph convolutional network (FPG-STGCN) to fast solve unit commitment (UC). Firstly, STGCN is tailored to parameterize UC. Then, few-shot physics-guided learning scheme is proposed. It exploits few typical UC solutions yielded via commercial optimizer to escape from local minimum, and leverages the augmented Lagrangian method for constraint satisfaction. To further enable both feasibility and continuous relaxation for integers in learning process, straight-through estimator for Tanh-Sign composition is proposed to fully differentiate the mixed integer solution space. Case study on the IEEE benchmark justifies that, our method bests mainstream learning ways on UC feasibility, and surpasses traditional solver on efficiency.
\end{abstract}

\vspace{-2em}

\section{Introduction}
\label{sec:introduction}
\IEEEPARstart{U}{nit} commitment (UC) is an exemplary NP-hard problem due to its mixed integer programming (MIP) formulation. Development of renewable high-penetrated power grids rises more complexity, since discrete space of UC are enlarged with tremendous integer-valued flexible resources, piecewise linearization of nonlinearities, discrete conditional operational rules, etc. These barriers have urged faster solvers for UC.

Several practical solvers have been developed \cite{UC_r1}. However, the sophisticated problem-specific heuristics or recursive branching incur time and resource-consuming UC. With that, data-driven methods have been thriving in. Existing roadmaps generally include three kinds, i.e., mimicking heuristics \cite{UC_r1}, enhancing substructures of traditional method \cite{UC_r3}, and learning mapping between given conditions and solutions \cite{UC_r2}, \cite{UC_r4}. To exemplify, \cite{UC_r1} presented a neural diving heuristic to assign predictions for partial variables and help quickly roll-out sub-MIPs. While in \cite{UC_r2}, supervised learning-to-branch method was reported to accelerate branching of branch-and-bound (B\&B). Yet, several flaws limit applications of these former two categories in UC. A erroneous prediction could result in no improvement in the B\&B search, and incur overtime to hit good dual gap and feasibility, which is unacceptable in power systems since one ISO must prioritize a feasible UC to ensure stable operation. Therefore, the third kind attracts more attention in power systems, since it can fast approach a feasible near-optimum. While supervised learning (SL) \cite{UC_r3}, and reinforcement learning (RL) \cite{UC_r4} are mainstreams in such framework. Albeit they do have ameliorated the solving procedure for UC, some critical deficiencies still persist. The principal limitation is that they demand external safety filter or corrector to ensure feasibility \cite{UC_r5}. This may compromise the coherence of the training graph, thereby influencing the learning stability, and also rise extra elapsed time. On the other hand, RL is also too hyperparameter-sensitive to settle generalized policies across non-stationary power system environments. Additionally, both SL and RL still depend on NP-hard discrete space search. In specific, SL needs MIP solvers to generate rich high-quality UC solutions to learn, and RL spans policy over discrete spaces with exponential complexity.


On the above obstacles, this letter seeks a new end-to-end learning-to-solve UC scheme that holds the following properties: 1) low-demand on labeled UC samples, 2) one-step decision towards feasible near-optimum, and 3) fully differentiable learning workflow to mitigate NP-hard while uphold UC constraints. To this end, we firstly apply a STGCN to parameterize UC, where binaries are represented by Tanh-Sign composition. Then, inspired by physics-guided learning (PGL) justified in continuous optimization task \cite{UC_r6}, augmented Lagrangian is applied to regularize the parameterized UC into hybrid loss function to enforce constraints. Straight-through estimator (STE) for Tanh-Sign composition is finally presented to differentiate UC along with integer feasibility.

\vspace{-1em}
\section{Unit Commitment Problem Statement}
Let $\Omega$, $\Omega_G$, and $\Omega_R$ represent the set of buses, synchronous generators, and renewable farms, respectively. The UC problem is generally formulated as follows ($t \in T$)
\vspace{-0.5em}
\begin{eqnarray}  
    \label{eq_1}
    \min \sum_{t \in T} \! \left\{ \sum_{i \in \Omega_G} \! [f\!(\!P_{G,i,t}) \! + \! C (S_{i,t})] \!+\!\sum_{i \in \Omega_R}\! ( P_{R,i,t}^F \! - \! P_{R,i, t}\!) \! \right \}
\end{eqnarray}
\vspace{-1em}
\begin{eqnarray}  
    \label{eq_2_20231018}
    \rm \mathbf s. \rm \mathbf t.  && \sum_{g \in \Omega_G \cap i}  
S_{g,t}P_{\emph G,\emph g,\emph t} + \sum_{r \in \Omega_R \cap i}  P_{\emph R,\emph r,\emph t} -  P_{\emph D,\emph i,\emph t} = P_{\emph i,\emph t},\nonumber \\
&&P_{i,t} = \sum_{j \in i} B_{ij}(\delta_{i,t} - \delta_{j,t}), \forall{i \in \Omega}
\end{eqnarray}
\vspace{-1.5em}
\begin{subequations}
\label{eq_5}
    \begin{eqnarray} 
    \label{eq_5a}
    \sum _{i 
    \in \Omega_G} \{ R_{G,i,t}^{u} = \min (S_{i,t} P_{G,i}^{max} - P_{G,i,t}, S_{i,t} r_{i}^{u}) \} \geq R_t^{u}
\end{eqnarray}
\vspace{-1.5em}
\begin{eqnarray} 
    \label{eq_5b}
    \sum _{i 
    \in \Omega_G} \{ R_{G,i,t}^{d} = \min (P_{G,i,t} - S_{i,t} P_{G,i}^{min}, S_{i,t} r_{i}^{d}) \} \geq R_t^{d}
\end{eqnarray}
\end{subequations}
\vspace{-1em}
\begin{equation}  
    \label{eq_6}
    - P_{ij}^{max} \le P_{ij,t} =  B_{ij}(\delta_{i,t} - \delta_{j,t}) \le P_{ij}^{max}, \forall {i, j \in \Omega}
    \end{equation}
\begin{equation}    
    \label{eq_7}
    S_{i,t} P_{G,i}^{min} \le P_{G,i,t}\le S_{i,t} P_{G,i}^{max}, \forall {i \in \Omega_G}
    \end{equation}
\begin{equation}    
    \label{eq_7s}
    0 \le P_{R,i,t}\le P_{R,i,t}^{F}, \forall  {i \in \Omega_R}
\end{equation}    
\begin{equation}  
    \label{eq_8}  
    P_{G,i,t-1} - r_i^d \le P_{G,i,t} \le P_{G,i,t-1} + r_i^u, \forall {i \in \Omega_G}
    \end{equation}
\vspace{-2ex}
\begin{subequations}    
\label{eq_9}
    \begin{equation}
    \label{eq_9a}
    (X_{i,t-1}^{on} - T_i^{on}) (S_{i,t-1} - S_{i,t}) \ge 0, \forall {i \in \Omega_G}
    \end{equation}
    \begin{equation}
        \label{eq_9b}
        (X_{i,t-1}^{off} - T_i^{off}) (S_{i,t} - S_{i,t-1}) \ge 0, \forall {i \in \Omega_G}
        \end{equation}
\end{subequations}    
where $T$ stands for the set of optimization periods. $P_{G,i,t}$ and $P_{R,i,t}$ are dispatch of generators and renewable at bus $i$ at period $t$. $P_{R,i,t}^F$ is the forecast of the $i$th renewable unit. $S_{i,t}$ represents on/off status of the generator $i$, where 1 specifies the startup and 0 otherwise. $P_{D,i,t}$ is the load of $i$th bus. $P_{ij,t}$ is the active power transfer from bus $i$ to bus $j$ through line $ij$. $R_t^{d}$ and $R_t^{u}$ are the required downward/upward reserve capacity for the whole system, usually set as $15\%\sim20\%$ of the total load. $ r_i^d$ and $ r_i^u$ are the down/up ramp rate of the generator $i$. $X_{i,t}^{on}$ and $X_{i,t}^{off}$ are continuous startup/shutdown time of the generator $i$, whilst $T_i^{on}$ and $T_i^{off}$ are the minimum duration for which the generator is required to sustain on/off status. $j \in k$ denotes that bus $j$ is connected to bus $k$. (\ref{eq_1}) is objective function that minimizes the operating cost of conventional units and renewable curtailment. (\ref{eq_5}) enforces the reserve capacity to guarantee intraday power balance under disturbances. (\ref{eq_6}) secures the operations of transmission lines.

\section{Proposed Learning-to-solve Scheme for UC}
\subsection{STGCN-based parameterization for UC solutions}
We consider STGCN with two crucial adaptations. First and foremost, the incorporation of constraints (2) and (4) delineates the spatiality of the UC model. To effectively master this spatial aspect, GCN becomes our prior. Secondly, constraints (3), (6), and (7) feature temporality of UC. This necessities the ability to master the graph-structured time-series. To place the STGCN, a graph mirrored from the power grid is built at the outset, denoted by $\mathcal G(\mathcal V,\mathcal E,\boldsymbol{A})$. $\mathcal{V}$ and $\mathcal{E}$ are the sets of nodes and edges, corresponding to that of buses and lines, respectively. $\boldsymbol{A}$ is the adjacency matrix. Laplace matrix is then defined as $\boldsymbol L=\boldsymbol D-\boldsymbol A$, where $\boldsymbol D$ is degree matrix, which is the diagonalization of the number of the neighbors near a node. A spatio-temporal convolutional block is further utilized \cite{r7-stgcn}. It allows the spatiotemporality to propagate across convolutions.

After that, we feed the forecasting profiles of nodal loads and renewable generation into the designed STGCN, and conduct the above propagation mechanism to yield the responses. By defining the STGCN responses as the desired decision variables, parameterization for UC is completed. The intact process can be mathematically depicted as follows:

\vspace{-1.5em}
\begin{subequations}
    \label{10}
    \begin{equation}
        \label{10a}
        \boldsymbol{H}(\boldsymbol\cdot)^{(0)} = \boldsymbol{x} =[\boldsymbol{x}_1, \cdots, \boldsymbol{x}_t, \cdots, \boldsymbol{x}_{|T|}]^{\top}, \forall t \in T
    \end{equation}
    \vspace{-1.5em}
    \begin{equation}
        \label{10b}
        \boldsymbol{x}_t = 
        \begin{bmatrix} 
        P_{D,1,t}, \cdots, P_{D,i,t}, \cdots, P_{D,N,t} \\
        P^{F}_{R,1,t}, \cdots, P^{F}_{R,i,t}, \cdots, P^{F}_{R,N,t}
        \end{bmatrix} ,\forall i \in \Omega
        \end{equation}
\end{subequations}

\vspace{-1.5em}
 \begin{equation}
\label{eq_10} 
\boldsymbol{H}(\boldsymbol\cdot)^{(l+1)} = \boldsymbol{\Gamma}^l_1 * \tau \sigma \! \left( \!\underbrace{\vphantom{\sum_{k=0}^{K-1} \boldsymbol\theta_k  \boldsymbol{T}\!_k\!(\!\boldsymbol{\tilde{L} }\!)}
 \sum_{k=0}^{K-1} \boldsymbol\theta_k  \boldsymbol{T}\!_k\!(\!\boldsymbol{\tilde{L} }\!)}_{\text{Spatial graph-conv}}\! \underbrace{\vphantom{\sum_{k=0}^{K-1} \boldsymbol\theta_k  \boldsymbol{T}\!_k\!(\!\boldsymbol{\tilde{L} }\!)}\left(\!\boldsymbol{\Gamma}^l_0 \!* \!\tau \boldsymbol{H}(\boldsymbol\cdot)^{\!(l)}\!\right)}_{\text{Temporal gated-conv}}\! \right)
\end{equation} 

 \vspace{-1.5em}
 \begin{eqnarray}
      \label{eq_13_20231018}
    \boldsymbol{\mathcal{R}} \! &=& \!\boldsymbol{H}  (\boldsymbol\cdot) ^{(L)} \! = \!   \boldsymbol{\mathcal{\phi}} ^{\boldsymbol{\theta}}  (\boldsymbol x, \boldsymbol A) =[\boldsymbol{P}_G, \boldsymbol{s} ,\boldsymbol{\delta} ]
 \vspace{-0.5em}
 \end{eqnarray}
where $\boldsymbol{x}$ is the input tensor and is distributed over the nodes of the graph $\mathcal{G}$. $\boldsymbol{\mathcal{R}}$ is the response tensor of the STGCN, and $[\boldsymbol{P}_G,\boldsymbol{\delta},\boldsymbol{s}]\!=\! \{[P_{G,i,t},\delta_{i,t},{s}_{i,t}]|i\in\Omega, t \in T\}$. Note that, there are nulls in $\boldsymbol{x}_t$ and $\boldsymbol{\mathcal{R}}$ due to mismatches among the set $\Omega$, $\Omega_{G}$, and $\Omega_{R}$. We use 0 as padding to solve this issue. $|T|$ is the cardinally of $T$. $L$ is the number of layers. $\boldsymbol{\Gamma}^l_0$,  $\boldsymbol{\Gamma}^l_1$ are the  upper and lower temporal kernel within the block of the layer $l$. $\boldsymbol{T}_k(\boldsymbol{\tilde{L}}) =\cos (k \cos^{-1}(\boldsymbol{\tilde{L}}))$ is Chebyshev polynomial respect to $\boldsymbol{\tilde{L}}$, and $\boldsymbol{\tilde{L}}=2\boldsymbol{L}/\lambda_{max} - \boldsymbol{I}$ , $\lambda_{max}$ denotes the largest eigenvalue of $\boldsymbol{L}$. $\boldsymbol\theta_k$ denotes the weight vector that connect convolution kernels. $*$ is the graph convolution operator. $K$ is the size of graph convolution kernel. $\boldsymbol{\mathcal{O}}=\mathcal{\phi}^{\boldsymbol\theta}(\boldsymbol{x}, \boldsymbol{A})$ in (11) is a simplified denotation for the nonlinear mapping between the given inputs $(\boldsymbol{x},\boldsymbol{A})$ and the desired outputs $\boldsymbol{\mathcal{O}}$ with functional of $\mathcal{\phi}^{\boldsymbol\theta}$ with parameter vector $\theta$.

Another significant component of UC-specified STGCN is the inference for integer-valued on/off status of generators. We use Tanh-Sign composition to transfer the intermediate output $\boldsymbol{s}$ into that, as shown below.

\vspace{-2em}
\begin{equation}
    \label{eq_14_20231018}
    \boldsymbol{S}\!=\! \{S_{i,t}|i\in\Omega, t \in T\},{S}_{i,t} = \text{sign}(\tilde{s}_{i,t})=
    \begin{cases}
    1, \ \textit{if} \ \tilde{s}_{i,t} > 0 \\
    0, \ \textit{if} \ \tilde{s}_{i,t} \le 0
    \end{cases}
\end{equation}
where $\tilde{s}_{i,t} = \text{tanh}(s_{i,t})$. Following this transfer process, we define the final UC decision by $\boldsymbol{\mathcal{O}} = \boldsymbol{\mathcal{\psi}} ^{\boldsymbol{\theta}}  (\boldsymbol{x}, \boldsymbol A)=[\boldsymbol{P}_G, \boldsymbol{S} ,\boldsymbol{\delta} ]$. 

\subsection{The FPG-STGCN design and its learning scheme}
We then cope with the aforementioned pivotal issues. Firstly, under the premise of low-demand for high-quality UC samples, we train a STGCN to enable one-step decision towards feasible near-optimum. To this end, the responses of the STGCN should approach the limited ground-truths available, while adhering to the hard constraints (2)$\sim$(8), regardless of whether the training data are labeled or not. Whereupon, the FPG-STGCN design is proposed.

In specific, a training set $\boldsymbol{\mathcal{D}}=(\boldsymbol{x}, \boldsymbol{A})$ with few labeled shots is considered, where we denote the labeled subset as $\hat{\boldsymbol{\mathcal{D}}}=[(\boldsymbol{x}, \boldsymbol{A}), \hat{\boldsymbol{\mathcal{O}}}]$ ($\hat{\boldsymbol{\mathcal{O}}}$ is the UC solutions gathered by commercial optimizer), and $\hat{\boldsymbol{\mathcal{D}}} \subset \boldsymbol{\mathcal{D}}$. In the context, the hat notation is used to tag the true outputs and their ingredients of $\hat{\boldsymbol{\mathcal{D}}}$. 

\emph{Remark}: the selection of which samples from the overall sample pool to label is vital for training convergence. This is because the added non-convex physical loss components can trap the FPG-STGCN in local optima. Therewith, the labeled samples should be able to cover the dominant patterns implicit in dataset. To do this, we cluster input samples and use traditional optimizer to yield UC solutions for the clustered centroids, forming the highly representative labeled subset. Upon the labeld few-shot, the supervised component can be settled ($\tilde{\boldsymbol{s}}=\{\tilde{s}_{i,t}|i\in\Omega, t \in T\}$):

\vspace{-1.5em}
\begin{align}
    \label{eq_16_20231113}
      \mathcal{L}_{\hat{\boldsymbol{\mathcal{D}}}} (\mathcal{\psi}^{\boldsymbol{\theta}} \!(\boldsymbol{x}, \! \boldsymbol{A}),  \hat{\boldsymbol{\mathcal{O}}}) = \left\Vert \boldsymbol{P}_G - \hat{\boldsymbol{P}}_G \right\Vert^2 + \left\Vert \boldsymbol{\delta} - \hat{\boldsymbol{\delta}} \right\Vert^2 \nonumber \\
      - \hat{\boldsymbol{S}}\text{log}(\frac{1}{2}+\tilde{\boldsymbol{s}}) - (1-\hat{\boldsymbol{S}})\text{log}(\frac{1}{2}-\tilde{\boldsymbol{s}})
\end{align}

To simplify the follow-up description, we rewrite the UC model. Particularly, let $f_{o}$, $f_{C}$, and $f_{B}$ stand for objective, constraints with only continuous variables, and binary-included constraints, respectively, such that the optimization variables are replaced by $\psi^{\boldsymbol{\boldsymbol{\theta}}}  (\boldsymbol{x}, \boldsymbol A)$ and the UC optimizes $\boldsymbol{\theta}$. The second loss is then derived from this model. Since non-convexity is involved, augmented Lagrangian is applied \cite{r9-ALM}, resulting in:

\vspace{-1.5em}
\begin{eqnarray}    
\label{eq_augL} &\mathcal{L}_{\boldsymbol{\mathcal{D}}} \left( \boldsymbol{\lambda},\!\boldsymbol{x},\!\mathcal{\psi}^{\boldsymbol{\theta}}(\boldsymbol{x},\!\boldsymbol{A}) \right)= {f}_{o}\!\left( \boldsymbol{x},\! \mathcal{\psi}^{\boldsymbol{\theta}}\!(\boldsymbol{x}, \!\boldsymbol A ) \right) \nonumber \\
&+ \boldsymbol\lambda_i \left[ \right. {f}_i\left( \boldsymbol{x}, \!\mathcal{\psi}^{\boldsymbol{\theta}}\!(\boldsymbol{x}, \boldsymbol A )\! \right)  \left. \right ]^{\top}  \!+\! \frac{\rho}{2} \sum_{i} \left\Vert {f}_i\left( \boldsymbol{x}, \mathcal{\psi}^{\boldsymbol{\theta}}(\boldsymbol{x}, \boldsymbol A ) \right) \right\Vert_2^2    
\end{eqnarray}
where $\rho > 0$, $i \in \{C, B\}$. $\boldsymbol{\lambda}_i$ are the tensors of dual variables. 
\vspace{-1.2em}

Adding $\mathcal{L}_{\hat{\boldsymbol{\mathcal{D}}}}$ and $\mathcal{L}_{\boldsymbol{\mathcal{D}}}$ get the final loss function. To train such FPG-STGCN, we exploit the duality, and do alternate update for dual variables and $\boldsymbol{\theta}$. Given the $\boldsymbol{\theta}$ and $\boldsymbol{\lambda}$ at the $k$th iteration, their $(k+1)$th updates are shown as (15):

\vspace{-1.5em}
\begin{subequations}
    \label{ascent}
    \begin{equation}    
    \label{ascento}    \boldsymbol{\theta}^{k+1} \!=\!\arg\min\limits_{\boldsymbol{\theta}} \mathcal{L}_{\hat{\boldsymbol{\mathcal{D}}}}(\mathcal{\psi}^{\boldsymbol{\theta}},\!\hat{\boldsymbol{\mathcal{O}}}) + \mathcal{L}_{\boldsymbol{\mathcal{D}}}(\boldsymbol{\lambda}^k,\!\boldsymbol{x},\! \mathcal{\psi}^{\boldsymbol{\theta}})
    \end{equation}
    \vspace{-1.5em}
    \begin{equation}
        \label{ascentmul}
        \lambda_i^{k+1} =\lambda_i^{k} + \rho \left | {f}_i \left( \mathcal{\psi}^{\boldsymbol{\theta}^{k+1}}(\boldsymbol{x}, \boldsymbol{A}) \right) \right | 
    \end{equation}
\end{subequations}
\vspace{-1.5em}

Notably, (\ref{ascento}) is realized upon the gradient descent:
 \vspace{-0.5em}
 \begin{eqnarray}
    \label{eq_17_20231016}
    \boldsymbol{\theta} \!&\leftarrow &\! \boldsymbol{\theta} + \frac{\partial \mathcal{L}_{\hat{\boldsymbol{\mathcal{D}}}}(\psi^{\boldsymbol{\theta}},\hat{\boldsymbol{\mathcal{O}}})}{\partial \boldsymbol{\theta}} +  \frac{\partial\mathcal{L}_{\boldsymbol{\mathcal{D}}}(\boldsymbol{\lambda},\boldsymbol{x},\mathcal{\psi}^{\boldsymbol{\theta}})}{\partial\boldsymbol{\theta}}  \nonumber \\
    &=& \boldsymbol{\theta} + \frac{\partial \mathcal{L}_{\hat{\boldsymbol{\mathcal{D}}}}(\psi^{\boldsymbol{\theta}},\hat{\boldsymbol{\mathcal{O}}})}{\partial \boldsymbol{\theta}} +\frac{\partial\mathcal{L}_{\boldsymbol{\mathcal{D}}}(\boldsymbol{\lambda},\boldsymbol{x},\mathcal{\psi}^{\boldsymbol{\theta}})}{\partial\psi^{\boldsymbol{\theta}}}\frac{\partial\psi^{\boldsymbol{\theta}}}{\partial\phi^{\boldsymbol{\theta}}}\frac{\partial\phi^{\boldsymbol{\theta}}}{\partial\boldsymbol{\theta}}
\end{eqnarray}
\vspace{-1.5em}

In (16), all terms except $\frac{\partial\psi^{\boldsymbol{\theta}}}{\partial\phi^{\boldsymbol{\theta}}}$ are continuously differentiable, and in greater detail, the Tanh-Sign composition in $\frac{\partial\boldsymbol{S}}{\partial\boldsymbol{s}}$ is non-differentiable. To build an fully differentiable training graph, STE is conducted. Unfolding $\frac{\partial\boldsymbol{S}}{\partial\boldsymbol{s}}$ via the chain rule, we have:

\vspace{-0.5em}
\begin{equation}
    \label{eq_18_20231113}
    \frac{\partial\boldsymbol{S}}{\partial\boldsymbol{s}} = \frac{\partial \text{sign}(\text{tanh}(\boldsymbol{s}))} {\partial\text{tanh}(\boldsymbol{s})}\frac{\partial\text{tanh}(\boldsymbol{s})}{\partial\boldsymbol{s}}
\end{equation}
\vspace{-1em}
\begin{align}
\label{eq_18_20231016}
     \begin{split}
     \frac{\partial \text{sign}(\text{tanh}(\boldsymbol{s}))} {\partial\text{tanh}(\boldsymbol{s})} \overset{\mathrm{STE}}{=}  \left \{
         \begin{array}{cc}         
         1,       & \text{tanh}(\boldsymbol{s}) \in [-0.5, 0.5] \\
         0,       & \text{otherwise}
         \end{array}
         \right.
     \end{split}
\end{align}

STE uses linear function to approximate the gradient of the Sign function at the zero point \cite{r8-yin2019understanding_STE}. It enables the FPG-STGCN with the ability to propagate gradients continuously across the non-differentiable $\frac{\partial \text{sign}(\text{tanh}(\boldsymbol{s}))} {\partial\text{tanh}(\boldsymbol{s})}$ during the training process.

\section{Case Study}
The modified IEEE 30-bus system is used to testify the proposed method, where wind farms are integrated at bus 6, 12, 10, 15 and 27 with the rated power of 60MW, 70MW, 55MW, 65MW, and 65MW, respectively, resulting in a 39.62$\%$ of wind penetration rate versus the total installed capacity. Renewable forecasts and load data are gathered upon historical collections of a region with a 5-minute period in one year. Several methods are introduced to establish our comparative studies: 1) M0: Gurobi solver upon the B\&B algorithm \cite{r_10}, 2) M1: a SL rival which imitates M0's outputs using STGCN, 3) M2: our method excluding the few-shot learning ingredient, 4) M3: A2C utilizing the same GCN as ours, and 5) our method.

\begin{figure}[!t]
\setlength{\abovecaptionskip}{-0.02cm}  
\setlength{\belowcaptionskip}{-0.02cm} 
\centering
\includegraphics[width=3.5in]{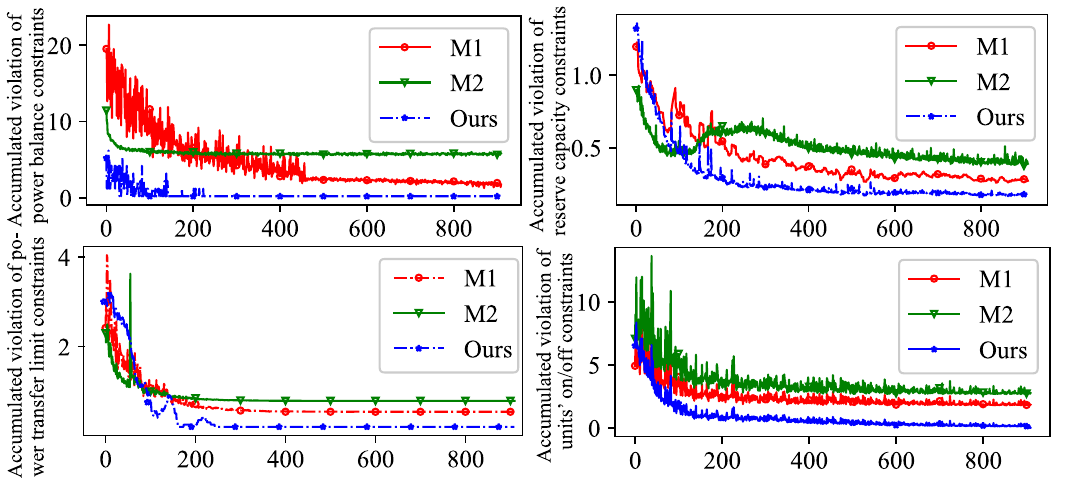}
\caption{Ablation test on constraint violation evolving with training epochs}
\vspace{-1ex}
\label{fig5}
\vspace{-2ex}
\end{figure}

To justify the necessity of each component in our FPG-STGCN, Fig. {\ref{fig5}} is provided. Fig. {\ref{fig5}} reveals that, M1 (SL) cannot eliminate constraint breach, as it merely mimics statistical patterns, completely disregarding physical constraints. Whilst M2 enforces constraints in learning UC decision, the intricate merger of continuous and discrete constraints, along with the STE module, has intensified the learning complexity and the risk of STGCN falling into local optima. Only when both the few-shot learning module and the physics-guided module coexist and mutually constrain each other, can STGCN learn effective solutions that satisfy the UC feasibility.
 
To further testify the performance of our method in online decision stage, several 
statistical indicators are carried out and the outcomes under 200 testing scenarios are collected in Tab. \ref{tab1}. As you can see from the Tab. \ref{tab1}, our method showed the smallest cost deviation (\$2.1±2.9), renewable curtailment (0.9±1.1MW) versus the M0 baseline. On the other hand, combined the results provided in Fig. 1 and that of the picked line overload frequency, the constraint satisfaction ability of our method is justified. These results indicate that, method can better approximate the optimal solutions provided by the solver than other learning algorithms. On the aspect of decision time, you can see that all learning-based methods significantly beat the traditional solver. While the learning-based approach may not yield a perfect UC solution akin to M0, it can serve as swift initialization for mainstream solvers with feasibility guarantee, thus expediting the UC solver procedure reliably.

\begin{table}
\scriptsize
\centering
\captionsetup{labelformat=empty}
\caption{Comparative outcomes regarding operating cost deviation $\text{E}_{Cost}$, renewable curtailment deviation versus M0 $\text{E}_{Curt.}$, online decision elapsed time, and line overload frequency under 200 testing scenarios (mean ± variance)}
\label{tab1}
\vspace{-3ex}
\begin{tblr}{
  width = \linewidth,
  row{1} = {c},
  row{2} = {c},
  row{3} = {c},
  row{4} = {c},
  row{5} = {c},
  row{6} = {c},
  row{7} = {c},
  hline{1-2,7} = {-}{},
}
\hline
 Method   & $E_{Cost}(\$)$ & $\text{E}_{Curt.}$(MW) & Time(s)& Freq.(\%) \\
 M0    &-    & -   & 5.247 $\pm$ 1.362&0 $\pm$ 0    \\  
 M1   &203.7  $\pm$ 196.4  & 12.1 $\pm$ 3.4 & 1.887 $\pm$ 0.305 &1.17 $\pm$ 0.51   \\
 M2   & 402.5 $\pm$ 358.3   &  14.2 $\pm$ 6.3 & 2.109 $\pm$ 0.307 &1.23 $\pm$ 0.53 \\
 M3  & 68.3 $\pm$ 47.5    & 3.2 $\pm$ 2.1     & 2.996 $\pm$ 0.371  &1.72 $\pm$ 0.53   \\
 Ours & 2.1 $\pm$ 2.9 & 0.8 $\pm$ 1.1   & 2.109 $\pm$ 0.308 & 0.018 $\pm$ 0.018   \\
\hline 
\end{tblr}

\vspace{1ex}
\footnotemark{ $ Freq. (\%) = n_{l}^{over}/n_{l}$, $n_{l}$ and $n_{l}^{over}$  indicate the number of branches and number of branches that exceed the power flow limit in a scheduling period T.}
\vspace{-2em}
\end{table}


\section{Conclusion}
This letter proposed a FPG-STGCN for learning-to-solve UC. It leverages the typical few-shot to help master the principal patterns, while introduces PGL to realize satisfaction for continuous and discrete constraints of UC. STE is finally utilized to maintain gradient conductivity of the training graph, even with added integer-valued constraints in the learning loss. Numerical studies justified the effectiveness and efficiency of the proposed method. Yet, we have not yet theoretically prove the respective proportions of the required sample sizes for few-shot and PGL components. This is our ongoing work.

\vspace{-1em}
\bibliographystyle{IEEEtran}
\bibliography{paper}



\end{document}